\documentclass[letter,twocolumn]{jpsj2}

\def\FIGxFIRST{
\begin{figure}[t]
 \begin{center}
  \includegraphics[width=\hsize]{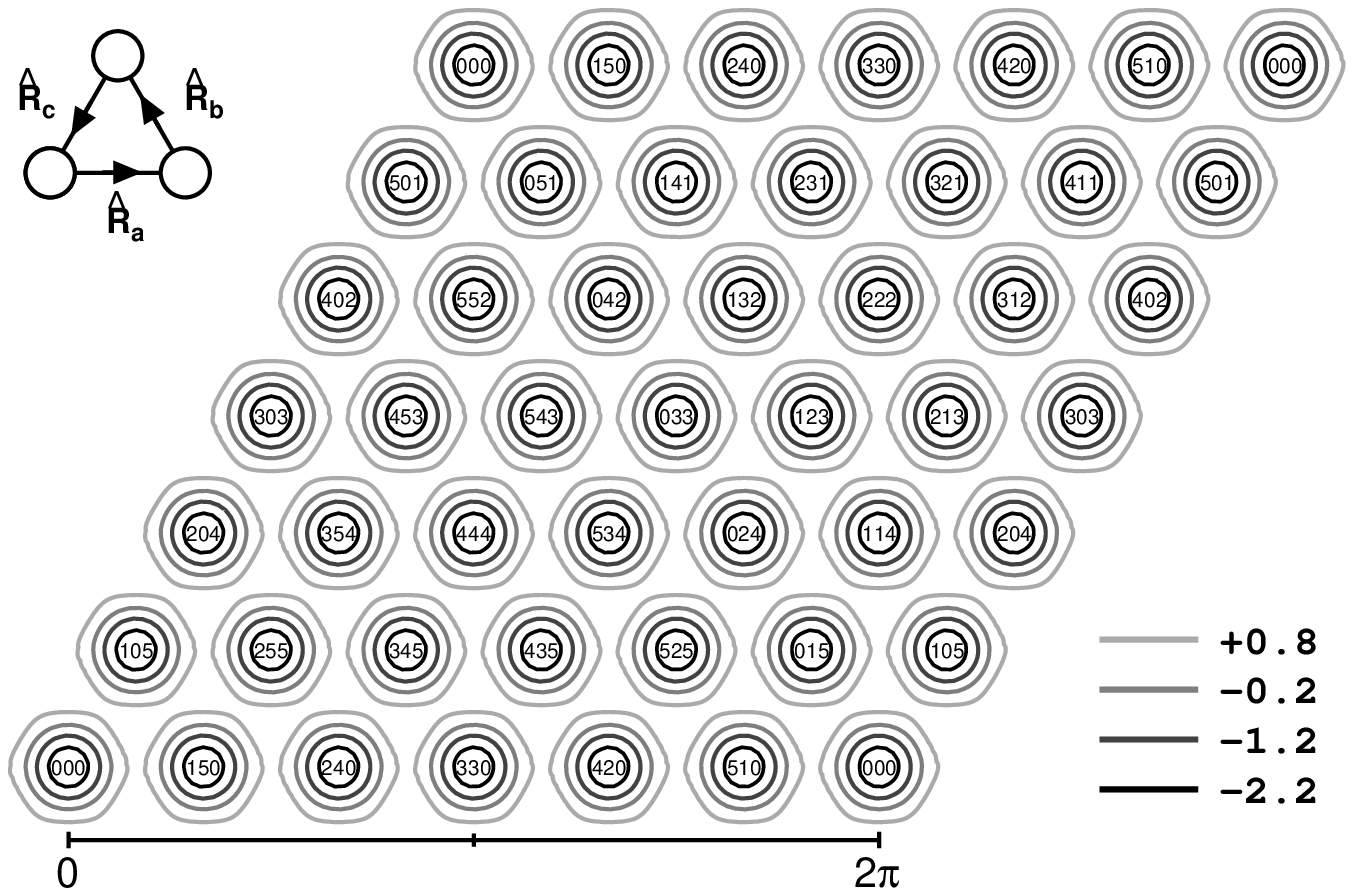}
 \end{center}
 \caption{
 The ideal-state graph ${\cal I}$ for $p=6$.
 Triplets of numbers exhibit 36 ordered states.
 The contour plot of the locking potential and the transformation eq.\
 (\ref{eq_gener_Phi}) (inset) are also given. 
 }
 \label{FIG1}
 \end{figure} }

\def\FIGxSECOND{
\begin{figure}[t]
\begin{center}
 \includegraphics[width=0.88\hsize]{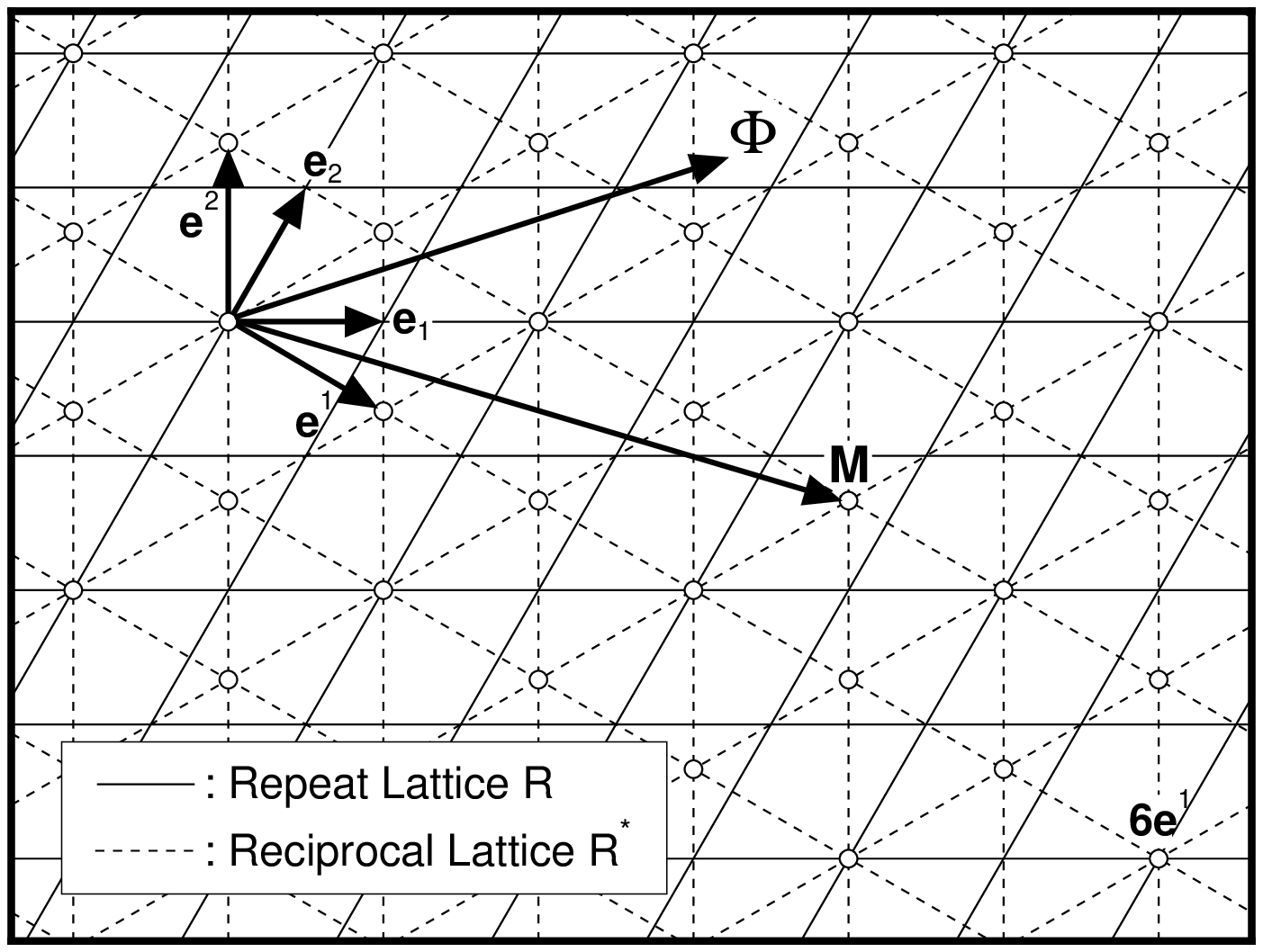}
\end{center}
 \caption{
 The repeat lattice ${\cal R}$ and its reciprocal lattice ${\cal R}^*$. 
 The fundermental lattice vectors are also given by arrows. 
 }
 \label{FIG2}
\end{figure} }

\def\FIGxTHIRD{
\begin{figure}[b]
\begin{center}
 \includegraphics[width=0.95\hsize]{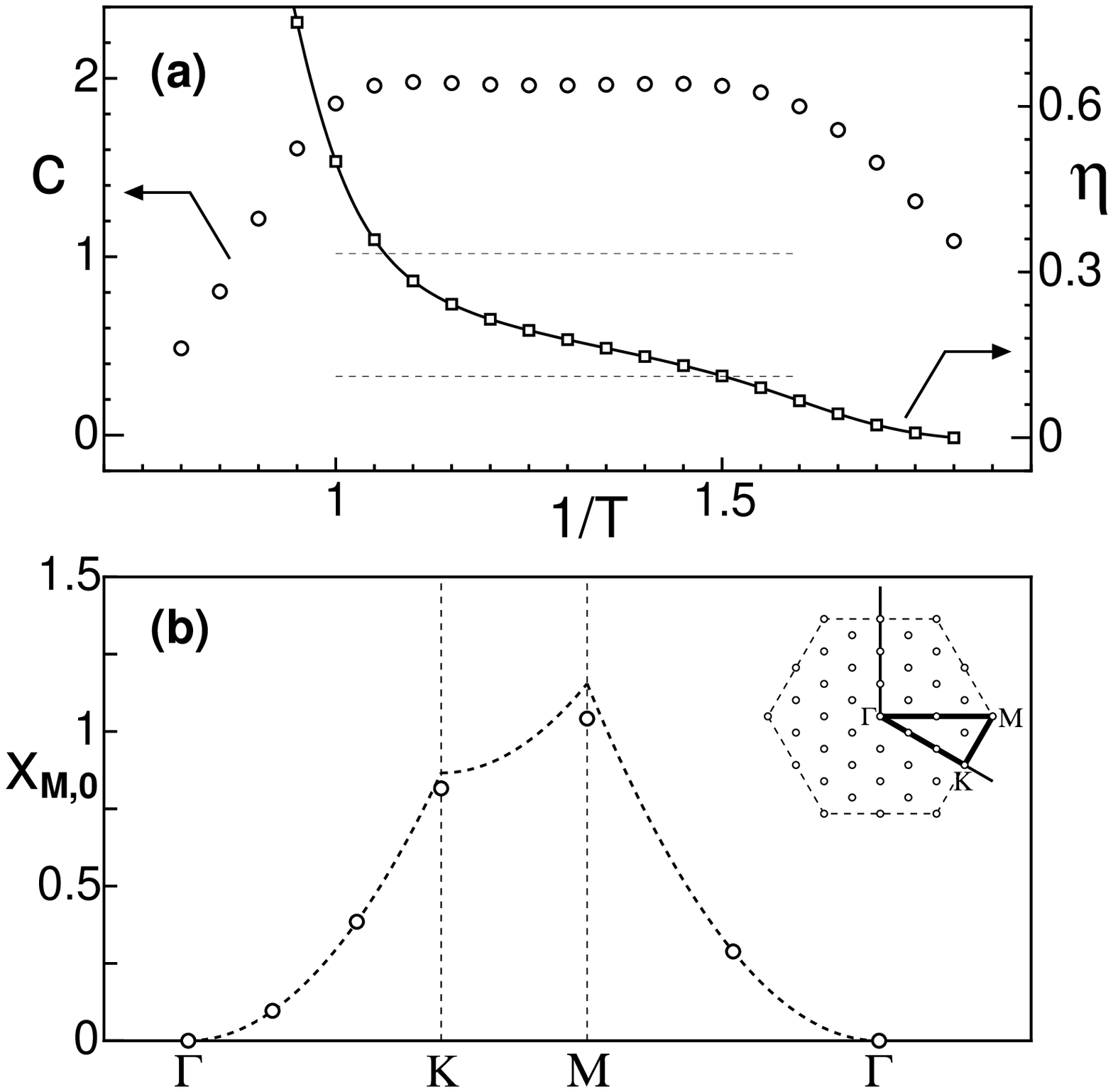}
\end{center}
 \caption{
 (a) $1/T$ vs. $c$ (circles) and $\eta$ (squares).
 (b) The ${\bf M}$ dependence of $x_{\bf M,0}$ at $1/T_{\rm SD}\simeq$1.252.
 Dotted curves exhibit
 $\|{\bf M}\|^2/2K$ with $K=4\sqrt3$ along the path in the inset.
 }
 \label{FIG3}
\end{figure} }

 \title{
 Effective Field Theory of Triangular-Lattice Three-Spin Interaction Model
 }

 \author{
 Hiromi \textsc{Otsuka}
 \thanks{E-mail address: otsuka@phys.metro-u.ac.jp}
 }

 \inst{
 Department of Physics, Tokyo Metropolitan University, Tokyo 192-0397, Japan
 }
 
 \recdate{\today}

 \abst{
 We discuss an effective field theory of a triangular-lattice three-spin
 interaction model defined by the ${\mathbb Z}_p$ variables.
 Based on the symmetry properties and the ideal-state graph concept, we
 show that the vector dual sine-Gordon model describes the long-distance
 properties for $p\ge5$; we then compare its predictions with the
 previous argument.
 To provide the evidences, we numerically analyze the eigenvalue
 structure of the transfer matrix for $p=6$, and we check the
 criticality with the central charge $c=2$ of the intermediate phase and
 the quantization condition of the vector charges.
 }

 \kword{
 triangular-lattice three-spin interaction model, 
 effective field theory,\\
 ideal-state graph,
 vector dual sine-Gordon model,
 two-dimensional melting
 }

\begin{document}
\maketitle

 The XY model consisting of the inner products of two neighboring planar
 spins with the ${\mathbb Z}_p$ symmetry-breaking field has been
 extensively researched; it provides the basic understanding of more
 complicated models and also of real materials.
 Especially, for the two-dimensional (2D) case, the effective theory for
 the long-distance behaviors was established based on the 2D Coulomb gas
 (CG) picture and the renormalization-group (RG) concepts, thereby
 enabling its complete
 understanding.\cite{Jose77}
 In this paper, we investigate its extension, i.e., the three-spin
 interaction model (TSIM) that was introduced a long time ago by Alcaraz
 et al.\cite{Alca82,Alca83}
 Suppose that $\langle k,l,m\rangle$ denotes three sites at the corners
 of each elementary plaquette of the triangular lattice $\Lambda$
 (consisting of interpenetrating sublattices $\Lambda_{\rm a}$,
 $\Lambda_{\rm b}$, and $\Lambda_{\rm c}$), then the following reduced
 Hamiltonian expresses a class of TSIM:
 \begin{eqnarray}
  {\cal H}
   =-\beta J\sum_{\langle k,l,m\rangle}
   \cos\left(\varphi_k+\varphi_l+\varphi_m\right). 
   \label{eq_Hamil}
 \end{eqnarray}
 Angle $\varphi_k=2\pi n_k/p$, $n_k\in[0,p-1]$ defines the
 ${\mathbb Z}_p$ variable.
 For $p=2$, eq.\ (\ref{eq_Hamil}) is the exactly solved Baxter-Wu model
 with three-spin-product
 interactions,\cite{Baxt73}
 but for larger $p$, it becomes puzzling in its expression in terms of
 the spin variables.
 However, for $p\ge5$ on which we will concentrate, its effective theory
 possesses a remarkably simple structure and can provide a unified
 understanding in a wide area of researches, including those on the 2D
 melting phenomena.
 Therefore, our aim is to formulate it based on a recent approach and to
 confirm its predictions quantitatively via numerical calculations. 

 We shall begin with the symmetry properties.
 Adding to the translations and space inversions, the model is invariant
 under the global spin rotations:
 $
 \varphi_k
   \mapsto \varphi_k
   +\sum_{\rho={\rm a,b,c}}\sum_{l\in\Lambda_\rho}(2\pi n_\rho/p)
   \delta_{k,l},
   $
 if the sublattice dependent numbers satisfy a condition
 $n_{\rm a}+n_{\rm b}+n_{\rm c}=0$
 (mod $p$).\cite{Alca82,Alca83}
 This symmetry operation---denoted as
 $(n_{\rm a}, n_{\rm b}, n_{\rm c})$---can
 be generated from two of the three fundamental operations
 with the following minimal spin rotations:
 \begin{equation}
  \hat R_{\rm a}\!: (1,p-1,0),~
  \hat R_{\rm b}\!: (0,1,p-1),~
  \hat R_{\rm c}\!: (p-1,0,1). 
  \label{eq_gener}
 \end{equation}
 Here, note that these commuting operations satisfy some relations,
 e.g.,
 $
 \hat R_{\rm a}^p=
 \hat R_{\rm c}\hat R_{\rm b}\hat R_{\rm a}=\hat1
 $ (identity operation).
 At a sufficiently low temperature, this symmetry can be broken due to the
 three-spin interaction, and one of the ordered states is stabilized.
 An obvious one is $\{\varphi_k\}=0$, and the others can be obtained by
 successively applying eq.\ (\ref{eq_gener}).
 Consequently, we can observe a $p^2$-fold degeneracy corresponding to
 the order of the
 group.\cite{Alca82,Alca83}

 Here, we shall mention that our strategy to formulate the effective
 field theory is based on the recent development by Kondev and
 Henley,\cite{Kond95,Kond96}
 where the critical ground states of various 2D classical systems have
 been treated in a unified way.
 So, to make it concrete, let us consider the structure of the so-called
 ideal-state graph $\cal I$ for the present problem. 
 We expect that, like the flat states in the interface models, the above
 $p^2$ ordered states should be specified by the locking points of a
 certain kind of field variable; thus, the dimension and the structure of
 ${\cal I}$ determine the number of components and the compactification
 region of the fields, respectively.
 By definition, each node of $\cal I$ represents one of the ordered
 states, and each link exhibits the neighboring two nodes connected by
 minimal spin rotations. 
 From the above-mentioned symmetry properties, there are two commuting
 generators, so the graph is located in a 2D space.
 Further, due to relations such as
 $\hat R_{\rm c}\hat R_{\rm b}\hat R_{\rm a}=\hat 1$, 
 the shortest loop of links should form the triangle. 
 These require $\cal I$ to be defined on the triangular lattice. 
 On one hand, for a given value of $p$, the periodicity conditions,
 e.g., $\hat R_{\rm a}^p=\hat 1$, define the unit cell of the so-called
 repeat lattice $\cal R$.
 From these, we can obtain ${\cal I}$. 
 In Fig.\ \ref{FIG1}, we give an example for $p=6$, where 36 ideal states
 are specified by the sublattice dependent numbers
 $n_{\rm a}$, $n_{\rm b}$, and $n_{\rm c}$
 (e.g., ``000'' at the corners represents $\{\varphi_k\}=0$).

 A two-component field theory is clarified to be relevant to the present
 model in its continuum limit; we shall write it as
 ${\bf \Phi}({\bf x})\in{\mathbb R}^2$
 (${\bf x}$ is the 2D real-space position vector).
 It then satisfies the periodicity conditions
 ${\bf \Phi}\equiv{\bf \Phi}+2\pi{\bf e}_\alpha$, 
 where $\{{\bf e}_\alpha\}~(\alpha=1,2)$ denote the normalized
 (non-orthogonal) fundamental lattice vectors of $\cal R$
 (see Fig.\ \ref{FIG2}).
 For the representations of the physical quantities in terms of ${\bf
 \Phi}$, it is also important to see the relationship between the
 discrete symmetry operations for spins and the field transformations,
 ${\bf \Phi}\mapsto{\bf \Phi}+\delta{\bf \Phi}$;
 especially, the correspondence of the minimal spin rotations eq.\
 (\ref{eq_gener}) to the following $\delta{\bf \Phi}$
 (see the inset of Fig.\ \ref{FIG1}):
 \begin{equation}
  \hat R_{\rm a}\!:  \frac{2\pi}{p}{\bf e}_1,~
  \hat R_{\rm b}\!: -\frac{2\pi}{p}{\bf e}_1+\frac{2\pi}{p}{\bf e}_2,~
  \hat R_{\rm c}\!: -\frac{2\pi}{p}{\bf e}_2,
  \label{eq_gener_Phi}
 \end{equation}
 which retain relations, e.g.,
 $\hat R_{\rm c}\hat R_{\rm b}\hat R_{\rm a}=\hat 1$.
 Our next task is to derive the Lagrangian density by which the
 low-energy and long-distance behaviors can be captured. 
 For this purpose, we shall first focus on the low-temperature region, at
 which the relevant potential may perturb the kinetic energy part
 representing the spatial fluctuation of ${\bf \Phi}({\bf x})$.
 As we shall see, the latter is also responsible for the description of
 the intermediate critical
 phase.\cite{Alca82,Alca83}

 \FIGxFIRST

 \FIGxSECOND

 Due to the periodicity of ${\bf \Phi}$, the vector charge $\bf M$ is
 quantized to take the values on the reciprocal lattice of $\cal R$,
 ${\cal R}^*$ (Fig.\ \ref{FIG2}).
 Then, the local densities relating to the spin degrees of freedom may
 be expanded to the Fourier series by using the vertex operators as
 $
 Q({\bf \Phi})=
 \sum_{{\bf M}\in{\cal R^*}} h_{{\bf M}}
 {\rm e^{i{\bf M\cdot\Phi}}}
 $.
 While the inner-product form ensures a coordinate invariance in
 ${\mathbb R}^2$, we look into its expression by using their elements to
 get some familiarity.
 Writing the reciprocal lattice vectors as
 $\{{\bf e}^\alpha\}~(\alpha=1,2)$,
 the inner product between
 ${\bf M}=m_\alpha{\bf e}^\alpha$
 and 
 ${\bf \Phi}=\sqrt2\phi^\alpha{\bf e}_\alpha$
 becomes $\sqrt2m_\alpha\phi^\alpha$, where the covariant and
 contravariant elements satisfy
 $m_\alpha\in{\mathbb Z}$ and $\sqrt2\phi^\alpha\equiv\sqrt2\phi^\alpha+2\pi$,
 respectively (the summation convention is used and the factor $\sqrt2$
 is for later convenience),
 since $\{{\bf e}_\alpha\}$ and $\{{\bf e}^\alpha\}$ are dual
 (i.e., ${\bf e}_\alpha\cdot{\bf e}^\beta=\delta_\alpha^\beta$).
 Also, the norm of ${\bf M}$, for example, is given by
 $\|{\bf M}\|=\sqrt{m_\alpha m^\alpha}$,
 where the metric tensor is defined by
 $g^{\alpha\beta}\equiv{\bf e}^\alpha\cdot{\bf e}^\beta$
 and the contravariant elements by
 $m^\alpha\equiv g^{\alpha\beta} m_\beta$,
 as usual.\cite{Comm1}

 To obtain an explicit form of the locking potential, the following
 three issues need to be
 addressed:\cite{Kond95,Kond96}
 (i)
 Since the potential is a part of the Lagrangian density, it should be
 invariant under eq.\ (\ref{eq_gener_Phi}).
 (ii)
 It is sufficient to keep the most relevant terms in the RG sense. 
 Since the dimension of a vertex operator is proportional to the squared
 norm of its vector charge (see below),
 it is sufficient to keep those with the shortest ones.
 (iii)
 The point group symmetry of the triangular lattice for ${\cal I}$, which
 stems from the sublattice and the spin-rotational symmetries, should be
 taken into account.
 Consequently, these requirements can be realized as the expression
 $V({\bf \Phi})=h_p \sum_{\|{\bf M}\|=pa^*} {\rm e^{i{\bf M\cdot\Phi}}}$,
 where
 $a^*\equiv\sqrt{g^{11}}$ and $h_p$ are the lattice constant of
 ${\cal R}^*$ and the coupling constant, respectively.
 The summation is over the following six vectors:
 $\pm p{\bf e}^1$,
 $\pm p{\bf e}^2$, and
 $\pm p({\bf e}^1+{\bf e}^2)$.
 In Fig.\ \ref{FIG1}, we give the contour plot of $V({\bf \Phi})$ for
 $p=6$ and $h_6=1/2$; we observe that the points with the minimum
 value form the triangular lattice and each point corresponds to the ideal
 state.

 While the locking point specifies the ordered state, the spatial
 fluctuation of the fields becomes important with the increase of the
 temperature.
 For $p\ge5$, the critical intermediate phase is
 expected,\cite{Alca82,Alca83}
 and it must correspond to the roughing phase of an interface
 model.\cite{Comm2}
 Thus, we can safely introduce the free-boson Lagrangian density for the
 fluctuations: 
 \begin{equation}
  {\cal L}_{0}[{\bf \Phi}]
   =\frac{K}{2\pi} \partial_i\phi_\alpha \partial_i\phi^\alpha,
   \label{eq_L0}
 \end{equation}
 where
 $g_{\alpha\beta}\equiv{\bf e}_\alpha\cdot{\bf e}_\beta$ and 
 $\phi_\alpha\equiv g_{\alpha\beta} \phi^\beta$. 
 The Gaussian coupling $K$ plays the role of inverse temperature to
 control the stiffness of the interface.
 The summation over $i$ $(=x,y)$ specifying the Cartesian component of
 ${\bf x}$ in the basal 2D plain is also assumed.

 At this stage, ${\cal L}_{0}+{\cal L}_{1}$ with 
 \begin{equation}
  {\cal L}_{1}[{\bf \Phi}]
   =\frac{y_{p,0}}{2\pi \alpha_0^2}
   \sum_{\|{\bf M}\|=pa^*} {\rm e^{i{\bf M\cdot\Phi}}}
   \label{eq_L1}
 \end{equation}
 ($y_{p,0}\propto h_p$ and $\alpha_0$ is the cutoff of $\Lambda$) can
 describe the lower temperature transition to the ordered phase.
 On the other hand, for the transition to the disordered phase,
 we should next consider the discontinuity of ${\bf \Phi}$ by an amount of
 $2\pi{\bf N}$ (${\bf N}\in{\cal R}$), which becomes frequent with the
 increase of the temperature. 
 This topological defect is created by the vertex operator
 ${\rm e^{i{\bf N\cdot\Theta}}}$, where ${\bf \Theta}$ is the dual field
 of ${\bf \Phi}$ and is defined as
 $K\partial_i{\bf\Phi}=\epsilon_{ij}\partial_j{\bf\Theta}$
 ($\epsilon_{ij}$ is the antisymmetric symbol). 
 ${\cal L}_{0}$ is invariant under the transformation,
 $K\leftrightarrow1/K$ and ${\bf\Phi}\leftrightarrow{\bf\Theta}$,
 so the stiffness of the interface described by ${\bf\Theta}$ becomes
 larger with the increase of the temperature. 
 Therefore, the potential can become relevant to bring about a unique
 flat state corresponding to the disordered phase.
 If we write
 ${\bf\Theta}\equiv\sqrt2\theta_\alpha{\bf e}^\alpha$,
 then the covariant elements satisfy
 $\sqrt2\theta_\alpha=\sqrt2\theta_\alpha+2\pi$,
 because the vector charge ${\bf N}$ is in ${\cal R}$.
 To obtain the explicit form of the locking potential, we can repeat the
 same argument as before, and arrive at the following form: 
 \begin{equation}
  {\cal L}_{2}[{\bf\Theta}]
   =\frac{y_{0,1}}{2\pi \alpha_0^2}
   \sum_{\|{\bf N}\|=1} {\rm e^{i{\bf N\cdot\Theta}}}, 
   \label{eq_L2}
 \end{equation}
 which gives the extremum on the lattice points of ${\cal R}^*$.
 Consequently, we see that the vector dual sine-Gordon field theory
 ${\cal L}\equiv{\cal L}_0+{\cal L}_1+{\cal L}_2$
 provides an effective description of TSIM.
 In the remaining part of this letter, we shall check our prediction
 analytically and numerically.
 Here, we summarize the scaling dimensions of the operators on
 ${\cal L}_0$.
 The vertex operator with the electric and magnetic vector charges
 $({\bf M, N})$
 is defined as
 $O_{\bf M,N}\equiv{\rm e}^{{\rm i}({\bf M\cdot\Phi}+{\bf N\cdot\Theta})}$,
 which has the dimension 
 $x_{\bf M,N}\equiv\frac12(K^{-1}\|{\bf M}\|^2+K\|{\bf N}\|^2)$.
 This formula supports our treatment that the vertex operators with the
 shortest vector charges were kept in the effective theory.
 In addition, the bosonized expression of the spin degrees of freedom,
 $S_k\equiv{\rm e}^{{\rm i}\varphi_k}$,
 is also important.
 It only contains the electric charges and should reproduce
 eq.\ (\ref{eq_gener}) on applying eq.\ (\ref{eq_gener_Phi}).
 We can see that the sublattice dependent expression,
 $
 (S_{\rm a},S_{\rm b},S_{\rm c})
 =
 (O_{ {\bf e}^1+{\bf e}^2,{\bf 0}},
  O_{-{\bf e}^1          ,{\bf 0}},
  O_{-{\bf e}^2          ,{\bf 0}})
 $, 
 satisfies the requirement and reproduces the ordered-state spin
 configurations on ${\cal I}$.
 Furthermore, these exhibit that the condition to give nonvanishing spin
 correlations
 \cite{Alca82,Alca83}
 is reduced to that of the vector charge neutrality.\cite{Kost73}
 For instance, since the two-spin (three-spin) product
 $S_\rho({\bf x})S^*_\sigma({\bf y})$
 $[S_\rho({\bf x})S_\sigma({\bf y})S_\tau({\bf z})]$
 becomes neutral for $\rho=\sigma$ ($\rho\ne\sigma\ne\tau\ne\rho$),
 the average with respect to ${\cal L}_0$ takes a nonzero value
 only between the same sublattice (among different sublattices).

 Now, we are in a position to discuss the transitions.
 Although the free part ${\cal L}_0$ is perturbed by ${\cal L}_{1,2}$
 (dimensions are $x_{y_{p,0}}=2p^2/3K$ and $x_{y_{0,1}}=K/2$),
 these are both irrelevant for $p^2/3\ge K\ge 4$.
 To obtain the lower and higher temperature transition points (say
 $T_{\rm L,H}$), we should resort to the numerical calculations, but
 these are considered to correspond to the above terminuses.
 Thus, let us first focus on the region near $T_{\rm H}$ where
 ${\cal L}_1$ is irrelevant.
 Instead of performing the perturbative RG calculation
 ($\alpha_0\to \alpha_0{\rm e}^{{\rm d}l}$),
 we shall see that our problem can be related to the works by
 Halperin, Nelson, and Young,\cite{Halp78,NelsHalp79,Youn79,Nels78}
 where the defect-mediated 2D meltings were studied based on the
 Kosterlitz-Thouless (KT) RG argument
 \cite{Kost73}
 (the so-called KTHNY theory).
 In fact, Alcaraz et al., making use of the vector CG representation with
 the long-range interaction, pointed out the relevance to 2D melting
 via the dissociation of dislocation pairs without the angular force.
 While our Lagrangian density ${\cal L}_0+{\cal L}_2$ is in the local
 representation, we can also find its counterpart in
 ref.\ \citen{Nels78}.
 Therefore, we confirm their assertion, and we shall discuss the
 transition in detail.
 As the Burgers vector characterizes the dislocation for the
 triangular lattice, ${\bf N}$ in eq.\ (\ref{eq_L2}) takes the six
 vectors,
 $\pm {\bf e}_1$,
 $\pm {\bf e}_2$, and
 $\pm({\bf e}_1-{\bf e}_2)$,
 so the three-point function of the local density ${\cal L}_2$ provides
 a nonvanishing value owing to the satisfaction of the neutrality
 condition.
 This results in a nonzero operator-product-expansion (OPE) coefficient
 among them; it yields the $y_{0,1}^2(l)$ term in the
 $\beta$-function for the coupling constant $y_{0,1}(l)$ (the fugacity
 for dislocations), while that for $x(l)\equiv2-x_{y_{0,1}}(l)$
 basically remains
 unchanged.\cite{Houg80}
 Consequently, we obtain the KT-like flow diagram
 (see Fig.\ 8 in ref.\ \citen{Nels78}).
 For $T>T_{\rm H}$, the correlation length behaves as
 $\xi\propto{\rm exp}[C/(T-T_{\rm H})^{\frac25}]$
 characterizing the disordered phase, while for $T<T_{\rm H}$
 the exponent of the spin correlation function varies as
 $\eta\equiv2x_{S}=4/3K$ and $\eta_{\rm H}=1/3$ at $T_{\rm H}$
 (without corrections). 

 Second, we shall discuss the region near $T_{\rm L}$ where ${\cal L}_2$
 is irrelevant.
 While having seen the $p^2$ ordered states by ${\cal L}_1$, we can also
 find a similar type of transition
 in refs.\ \citen{Halp78} and \citen{NelsHalp79},
 where the transition from the ``floating solid'' to ``commensurate solid''
 was discussed for the case that the substrate periodic potential is
 commensurate with the adsorbate lattice.
 Indeed, they observed its similarity to the dislocation-mediated
 melting.
 For the present model, Alcaraz et al. obtained the RG equations based on
 the duality observed in the vector CG
 representation.\cite{Alca82,Alca83}
 On the other hand, we can also reproduce them and also the finite
 $\xi$ for $T<T_{\rm L}$ and $\eta_{\rm L}=4/p^2$ at $T_{\rm L}$, etc,
 based on our local representation and OPE coefficients.

 \FIGxTHIRD

 To confirm the effective theory, we shall provide the numerical
 calculation results here.
 We consider the system on $\Lambda$ with $M$ ($\to\infty$) rows of $L$
 (a multiple of 3) sites wrapped on the cylinder and define the transfer
 matrix connecting the next-nearest-neighbor rows.
 We denote its eigenvalues as
 $\lambda_q(L)$
 or their logarithms as
 $E_q(L)=-\frac12\ln|\lambda_q(L)|$
 ($q$ specifies a level).
 Then, the conformal invariance provides the expressions of the central
 charge $c$ and the scaling dimension $x_q$ in the critical systems as
 $E_{\rm g}(L)\simeq Lf-{\pi}c/{6L\zeta}$ 
 and
 $\Delta E_q(L)\simeq {2\pi}x_q/{L\zeta}$. 
 Here,
 $E_{\rm g}(L)$, $\Delta E_q(L)$ $[=E_q(L)-E_{\rm g}(L)]$, $\zeta$
 $(=2/\sqrt3)$, and $f$ correspond to the ground-state energy, an
 excitation gap, the geometric factor, and a free energy density,
 respectively.\cite{Card84,Blot86,Affl86}
 When performing the diagonalization calculations, we employ two of
 three spin rotations in eq.\ (\ref{eq_gener})
 [e.g., $(\hat R_{\rm a},\hat R_{\rm b})$]
 as well as the lattice translation and space inversion.
 This is because the matrix size can be reduced, and more
 importantly, discrete symmetries can specify lower-energy excitations.
 For instance, we can find the excitation level corresponding to the
 spin operator $S_{\rm a}({\bf x})$ in the sector with indexes
 $({\rm e}^{{\rm i}2\pi/p},1)$.

 In the following, we give the results for $p=6$, which are obtained
 using the data up to $L=9$.
 In Fig.\ \ref{FIG3}(a), we provide the $T$ dependence (in units of
 $J/k_{\rm B}$) of $c$ where the region keeping $c\simeq2$ can be
 recognized.
 From this plot, we roughly estimate the KT-like transition temperatures
 as
 $1/T_{\rm L}\simeq 1.5$
 and
 $1/T_{\rm H}\simeq 1.1$
 (see also refs.\ \citen{Alca82} and \citen{Alca83}).
 Next, we plot the $T$ dependence of $\eta$ along its second axis, which
 shows that $\eta$ is the increasing function of $T$ and takes close
 values to $\eta_{\rm L,H}=1/9$, 1/3 around the estimates of
 $T_{\rm L,H}$ (see dotted lines).
 In Fig.\ \ref{FIG3}(b), we give the scaling dimension $x_{\bf M,0}$ as
 a function of the electric vector charge ${\bf M}$ along the path
 depicted in the inset.
 As a representative for the critical region, we pick up the self-dual
 point by numerically observing the level crossing,
 $x_{y_{6,0}}=x_{y_{0,1}}$; we then perform the calculations at 
 $1/T_{\rm SD}\simeq1.252$. 
 From the figure, we can verify that the dimension depends only
 on the norm of the electric vector charge.
 Further, despite the treated $L$ being small---the size dependence is
 actually larger for those with longer vectors---the results (open
 circles) agree well with the theoretical formula (dotted curves), which
 quantitatively supports our above argument.

 To summarize, based on the ideal-state graph approach by Kondev and
 Henley, we have discussed the effective field theory for TSIM.
 Due to the symmetry properties, two kinds of vector charges,
 ${\bf M}$ and ${\bf N}$,
 take the values of the reciprocal and repeat lattice
 points, and then provide the descriptions of the phase transitions. 
 While the effective theory is given by the vector dual sine-Gordon
 model, we have seen its close relationship to the KTHNY theory, as
 pointed out by Alcaraz et al.
 We also performed numerical calculations to check some theoretical
 predictions based on the conformal invariance.

 Finally, we make two remarks:
 (i)
 For the determination of $T_{\rm L,H}$, we observed the deviation from
 $c=2$.
 This could provide rough estimations, but a more efficient criterion is
 desired.
 For the $c=1$ case, the universal amplitudes of logarithmic corrections
 are utilized for the analysis of the excitation spectrum, which then
 offers level-crossing conditions for the KT-transition
 points.\cite{Mats05,Otsu05a}
 For this issue, ${\cal L}$ provides the basic framework to perform the
 one-loop calculations. 
 (ii)
 Apart from the above restriction $p\ge5$, it is also interesting
 to see whether ${\cal L}$ can describe the transitions observed in TSIM
 with $p<5$.
 In particular, for $p=2$, the existence/nonexistence of the critical RG
 flow connecting ${\cal L}_0$ to the ${\mathbb Z}_2$ orbifold of a
 Gaussian model with $c=1$ (the critical fixed point of the Baxter-Wu
 model), due to the competing relevant perturbations ${\cal L}_{1,2}$,
 may be an important
 issue.\cite{Lech02}
 We will address these issues in the future.

 The author thanks
 K. Nomura
 for stimulating discussions. 
 This work was supported by Grants-in-Aid from
 the Japan Society for the Promotion of Science. 

 \newcommand{\AxS}[1]{#1:}
 \newcommand{\AxD}[2]{#1 and #2:}
 \newcommand{\AxT}[3]{#1, #2, and #3:}
 \newcommand{\AxQ}[4]{#1, #2, #3, and #4:}
 \newcommand{\REF }[4]{#1 {\bf #2} (#4) #3}
 \newcommand{\PRL }[3]{\REF{Phys. Rev. Lett.\   }{#1}{#2}{#3}}
 \newcommand{\PRB }[3]{\REF{Phys. Rev.\        B}{#1}{#2}{#3}}
 \newcommand{\PRE }[3]{\REF{Phys. Rev.\        E}{#1}{#2}{#3}}
 \newcommand{\NPB }[3]{\REF{Nucl. Phys.\       B}{#1}{#2}{#3}}
 \newcommand{\JPA }[3]{\REF{J. Phys.\ A}{#1}{#2}{#3}}
 \newcommand{\JPC }[3]{\REF{J. Phys.\ C: Solid State Phys.}{#1}{#2}{#3}}


\begin{thebibliography}{99}
\bibitem{Comm2}
 Unlike the critical ground states, the $y_{0,1}$ term exists for our
 problem (see text); thus, the exact mapping to an interface model is
 absent.
 However, since it is irrelevant at low temperatures, we can drop it in
 the RG sense.
\end{thebibliography}
\end{document}